\documentclass[a4paper,fleqn,usenatbib]{mnras}

\usepackage{newtxtext,newtxmath}

\usepackage[T1]{fontenc}
\usepackage{ae,aecompl}

\usepackage{graphicx}	
\usepackage{amsmath}	
\usepackage{amssymb}	
\usepackage{soul}             
\usepackage{subfig}

\title[Reconnection Powered Blazar Flares from SSC]{The Feasibility of Magnetic Reconnection Powered Blazar Flares from Synchrotron Self-Compton Emission}

\author[P. J. Morris et al.]{
Paul J. Morris,\thanks{E-mail: paul.morris@physics.ox.ac.uk (PJM)}
William J. Potter, 
and Garret Cotter
\\
Oxford Astrophysics. Denys Wilkinson Building, Keble Road, Oxford, OX1 3RH, United Kingdom\\
}

\date{Accepted 2019 March 20. Received 2019 February 21; in original form 2018 October 9}

\pubyear{2018}

\begin{document}
\label{firstpage}
\pagerange{\pageref{firstpage}--\pageref{lastpage}}
\maketitle

\begin{abstract}

Order of magnitude variability has been observed in the blazar sub-class of Active Galactic Nuclei on minute timescales. These high-energy flares are often difficult to explain with shock acceleration models due to the small size of the inferred emitting region, with recent particle-in-cell (PIC) simulations showing that magnetic reconnection is a promising alternative mechanism. Here, we present a macroscopic emission model physically motivated by PIC simulations, where the energy for particle acceleration originates from the reconnecting magnetic field. We track the radial growth and relative velocity of a reconnecting plasmoid, modelling particle acceleration and radiative losses from synchrotron and synchrotron self-Compton (SSC) emission. To test the viability of magnetic reconnection as the mechanism behind rapid blazar flares we simultaneously fit our model to the observed light-curve and SED from the 2016 TeV flare of BL Lacertae. We find generally that, without considering external photons, reconnecting plasmoids are unable to produce Compton-dominant TeV flares and so cannot reproduce the observations due to overproduction of synchrotron emission. Additionally, problematically large plasmoids, comparable in size to the entire jet radius, are required to emit sufficient SSC gamma-rays to be observable. However, our plasmoid model can reproduce the rapid TeV lightcurve of the flare, demonstrating that reconnection is able to produce rapid, powerful TeV flares on observed timescales. We conclude that while reconnection can produce SSC flares on the correct timescales, the primary source of TeV emission cannot be SSC and the size of plasmoids required may be implausibly large.

\end{abstract}




\section{Introduction}

Blazars are the most luminous sub-class of Active Galactic Nuclei (AGN) \citep{1995PASP..107..803U} being those which have one of their relativistic plasma jets pointed towards us. This results in Doppler boosting which causes substantial amplification of the emitted flux from the jet, to the extent that it can dominate over any emission from the accretion disc or host galaxy. 

Blazars have a characteristic double humped continuum emission spectrum, spanning the electromagnetic spectrum from radio frequencies to very high energy (VHE) TeV $\gamma$-rays. The lower energy hump is caused by relativistic electrons and positrons undergoing helical motion in the magnetic field of the jet and emitting synchrotron radiation. In leptonic models, \citep[e.g.][]{1997A&A...320...19M,1998A&A...333..452K,1999MNRAS.306..551C,2000ApJ...536..729L,2002ApJ...581..127B} the higher energy bump is interpreted as inverse Compton (IC) emission \citep[e.g.][]{Longair:2011} where the relativistic electrons up-scatter the synchrotron seed photons (synchrotron self-Compton, SSC) or photons external to the jet to higher energies. Other models assume an additional hadronic component of the jet plasma \citep[e.g.][]{1993A&A...269...67M,1997ApJ...478L...5D,2002APh....17..347S,2013ApJ...768...54B}. These models include collisions between relativistic protons, causing the production of neutral pions, which then decay into a photon pair and provide the high energy blazar emission. Although these models have provided some good fits to the data, they often require a total power in relativistic protons many orders of magnitude higher than the observed bolometric power of the relativistic jet or the estimated Eddington luminosity \mbox{\citep{2013ApJ...768...54B,2015MNRAS.450L..21Z}} which is hard to justify.

Many models with varying jet geometry have been postulated to explain the broadband SEDs. Commonly, the dominant emission region within the jet is approximated as a compact spherical region \citep[e.g.][]{1996ApJ...461..657B,2002ApJ...581..127B,2007A&A...476.1151T}. Other models assume the jet has conical geometry such as the pioneering paper by \citet{1979ApJ...232...34B} and more recent work by \citet{1980ApJ...235..386M,1981ApJ...243..700K,1985A&A...146..204G} or the relativistic fluid models of \citet{PC:2012,PC:2013}. The large scale conical jet geometry is supported by VLBI observations from \citet{2013ApJ...775..118N} and \citet{2018ApJ...860..141H}.

Though the aforementioned models have in general been very successful in explaining quiescent emission in relativistic jets, problems still remain when adding time variability. One possible and popular explanation is that acceleration occurs via internal shocks within the jet, caused by variations in the bulk Lorentz factor of the jet leading to collisions of different emitting regions. Shock acceleration is a well established mechanism for particle acceleration \citep{1978MNRAS.182..147B,1987PhR...154....1B,2001MNRAS.325.1559S}, yet light travel time arguments prohibit large scale shocks from powering the most rapid flares \citep{Aharonian:2007}. Shock acceleration also requires particles cross the shock front multiple times to be accelerated up to the highest energies, which is difficult to achieve if the shock is magnetised and relativistic \citep{2015MNRAS.450..183S,2018MNRAS.473.2364B}. These are likely conditions in the case of astrophysical jets. 

TeV flares have been observed to occur on minute time scales for several active galaxies \citep{Aharonian:2007,Albert:2007,IC310:TeV,2018ApJ...856...95A}. Such observations pose significant theoretical challenges, since light travel time arguments imply that the radius, $R$, of the emitting region for a flare of duration $t_{\rm{var}}$ from a source at redshift $z$ cannot exceed \citep[e.g.][]{2007ApJ...671L..29L},
\begin{equation}
R \leq \frac{c t_{\rm{var}} \delta}{(1+z)}.
\label{size}
\end{equation}
For example, taking a $z=1$ source exhibiting a $10~$minute flare and with a typical Doppler factor of $\delta \approx 10$ implies a relatively compact region size of $R \leq 9 \times 10^{11}~$m.

\begin{figure}
	\includegraphics[width=0.45\textwidth]{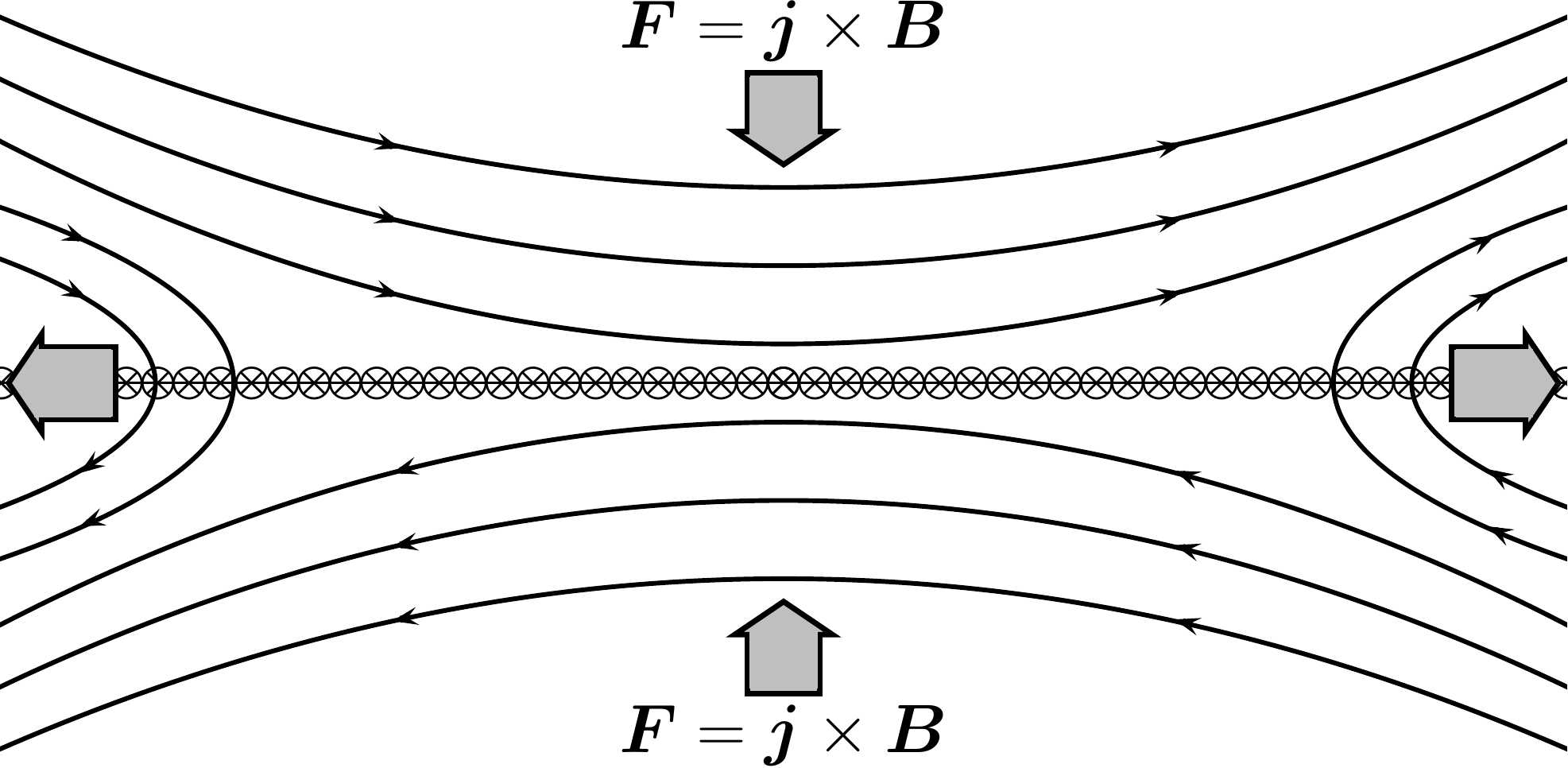}
    \caption{Schematic illustrating magnetic reconnection. Oppositely orientated magnetic field lines converge, creating a current sheet between them. In this figure, the direction of the current density ${\boldsymbol j}$ is into the plane of the paper, such that the direction of the Lorentz force, ${\boldsymbol F} = {\boldsymbol j} \times {\boldsymbol B}$ is indicated by the grey arrows. This force acts on the field lines forcing them into the reconnection region. It can be seen that the direction of ${\boldsymbol j} \times {\boldsymbol B}$ changes when the magnetic field lines reconnect, helping to expel material from the reconnection region. This creates a gas pressure gradient allowing fresh plasma to flow into the reconnection region and be accelerated outwards at the expense of magnetic energy.}
    \label{fig:MR}
\end{figure}


The necessity for a compact emitting region has led to a rise in popularity of magnetic reconnection as a mechanism which may be able to provide the required particle acceleration and compact emission regions needed for TeV flaring. The original framework for magnetic reconnection was developed by \citet{1958IAUS....6..123S} and \citet{1957JGR....62..509P}, yet this so called Sweet-Parker model is too slow to reproduce the observed timescale in solar flares \citep{2016imm..book.....G}. Analytic results have indicated that faster reconnection may be realised if the current sheet is subjected to instabilities which cause it to fragment into a chain of magnetic islands, or plasmoids \citep{2007PhPl...14j0703L,2010PhRvL.105w5002U}. Subsequent research utilising Particle-in-cell (PIC) simulations have verified that these effects can indeed cause more rapid reconnection \citep[e.g][]{2014ApJ...783L..21S}. The plasmoids which form from the initial current sheet have a range of properties from being large and mildly relativistic to small and ultra-relativistic \citep{2016MNRAS.462.3325P}. Collisions and subsequent merging of these plasmoids can result in particle acceleration with spectral indices between 1 and 2 \citep{2014ApJ...783L..21S,2014PhRvL.113o5005G,2018ApJ...862...80B}. Whilst the exact attribution of the acceleration mechanisms remains unclear, it has been postulated that the mean energy gain of the particles is approximately first order and therefore analogous to the exponential increase in energy provided by shock models \citep{2005A&A...441..845D,2014PhRvL.113o5005G,2015ASSL..407..373D, 2016MNRAS.463.4331D}. Other hypotheses include those where the acceleration is dominated by the reconnection electric field ($\boldsymbol{E} \approx - \boldsymbol{v} \times \boldsymbol{B}$) \citep[e.g.][]{2015SSRv..191..545K,2017ApJ...849...35I}. Generally, it seems likely that there is some contribution to particle acceleration from both mechanisms. 

\begin{figure*}
	\includegraphics[width=15cm]{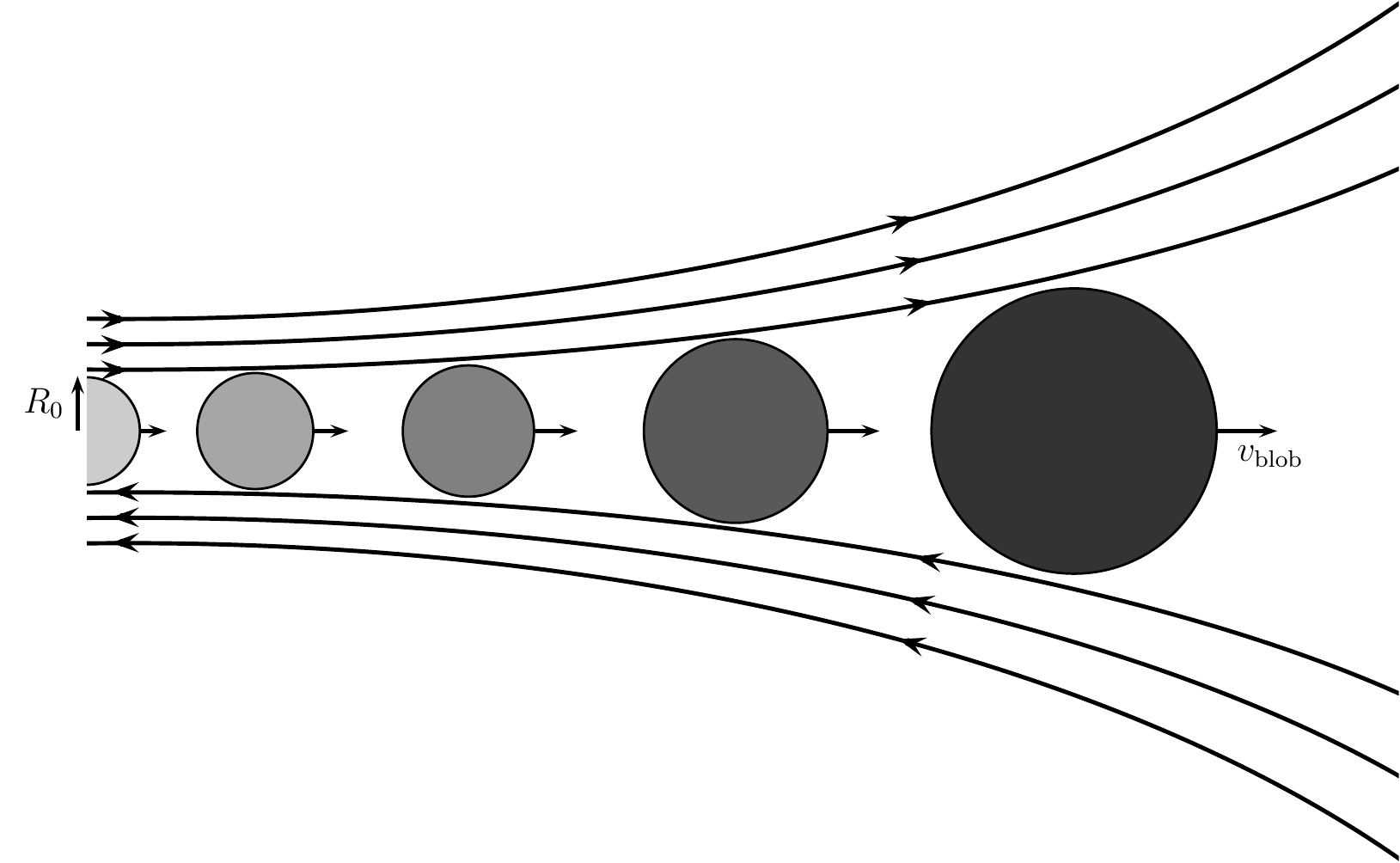}
    \caption{Schematic of the model. The diagram shows the evolution of a single plasmoid as it travels through the reconnection layer between regions of opposing magnetic field lines. Darker shading represents later times. The radius of the plasmoid steadily increases over time, as we assume it continuously merges with like plasmoids in the reconnection layer. In our model, we parameterise the space and relative velocity between plasmoids using the parameter $f$ and assume the acceleration timescale is the merge timescale, which is defined in Eqn. \ref{eq:tm}. The plasmoid velocity also varies with time as it is accelerated from its point of origin in the reconnection layer \citep[see][]{2016MNRAS.462...48S}. The total energy of the plasma is conserved in the reconnection process, so the energy for particle acceleration comes from the magnetic energy density, causing the magnetic field decreases with time. We assume the plasmoid remains in the reconnection layer until the energy initially stored in the reconnecting field has been transferred to particles and radiation. This defines the final plasmoid radius, $R_{\rm{f}}$, and the total distance travelled by the blob in the reconnection layer, $L$. }
    \label{modelFig}
\end{figure*}

Attempts to scale up these PIC simulations to macroscopic scales have been made, with the jets-in-a-jet model of \citet{GUB:2009} containing blobs moving relativistically within a jet indicating that it is possible to produce flares on the rapid observed timescales. Further work by \citep{NGBUS:2011} investigating the radiative properties of these mini-jets found an overproduction of X-ray emission. \citet{2013MNRAS.431..355G} discuss characteristic flaring profiles from models of plasmoid driven reconnection where fast flares from large ``monster" plasmoids are superimposed onto emission from the reconnection current layer. These results have been initially promising, and so warrant further investigation. 

Whilst PIC simulations have demonstrated some initial promise with regards to the feasibility of reconnection powered blazar flares, their computationally expensive nature restricts them to small spatial scales, as the plasma skin depth is resolved \citep[e.g.][]{2014ApJ...783L..21S,2016MNRAS.462...48S}. They are therefore constrained to smaller regions relative to the size of astrophysical jets required to model blazar flaring. In this paper, we present a macroscopic emission model, which is physically motivated by the results of PIC simulations. Our model computes the time evolution of a reconnecting plasmoid whose radius and velocity evolve as it travels through the reconnection layer. We calculate the acceleration and radiative losses of the electrons and compare our results to SED data for BL Lacertae, and output flaring profiles at TeV energies. Here, particle acceleration is computed for the constituent particle population contained within a growing and accelerating plasmoid in a reconnection layer, offering an improvement on previous models. We discuss the required physical parameters, and their implications for the feasibility of reconnection powered blazar flares. 

The structure of the paper is as follows: In Section \ref{model}, we begin by outlining the physical motivation of our reconnection model and discuss the particle acceleration process and computation of the radiative losses. In Section \ref{sec:resdisc} we discuss general properties of the high energy flaring profiles and SEDs produced by our model before presenting the fit of our model to the 2016 TeV flare of BL Lacertae \citep{2018ApJ...856...95A}. We then discuss the implications of our results with respect to PIC simulations, and constrain the location of the reconnection layer within the jet plasma before stating our conclusions in Section \ref{sec:conc}.

\section{Reconnection Model} \label{model}

We aim to model the emission across the entire electromagnetic spectrum for a radiatively emitting reconnecting plasmoid, paying particular attention to the high energy gamma-ray emission. As reconnection has been postulated to explain TeV flares in blazars, we output flaring profiles at these energies from our model to evaluate the emission timescales. In this section we outline the physical motivation for our model before summarising the particle acceleration and radiative loss computation in our code.

We assume a leptonic jet consisting entirely of electrons and positrons (hereafter electrons) and consequently model the acceleration and radiative losses for electrons that have entered the reconnection layer. The model begins with a single relativistic plasmoid, which is approximated as an initially highly magnetised sphere \citep[see][]{2016MNRAS.462...48S}. The total energy density in the rest frame of this plasmoid is initially given by,

\begin{equation}
U_{\rm{TOT}} = U_{\rm{B,0}} + U_{\rm{e,0}} = \left( \sigma_0 + 1 \right) U_{\rm{e,0}},
\label{eq:Ubal}
\end{equation}
where the magnetic energy density is $U_{\rm{B}} = B^2/2 \mu_0$ for magnetic field strength $B$. The ratio of magnetic to particle energy densities, $\sigma = U_{\rm{B}}/U_{\rm{e}}$, defines the magnetisation and the subscript 0 denotes initial values. We consider the total magnetic field to constitute the reconnecting magnetic field, $B_{\rm{rec}}$, which is depleted as magnetic energy is transferred to particles, and a guide field, $B_{\rm{g}}$, which remains constant and does not reconnect. The total magnetic field in the model is given by the sum of the guide and reconnecting field components. Throughout the computation, magnetic energy is converted to particle energy. This causes the value of $B_{\rm{rec}}$ to decrease with time as the magnetic energy is transferred to particles.  $B_{\rm{rec}}$, $B_{\rm{g}}$ and $\sigma_0$ are initialised as free parameters. We assume that the emitting plasmoid remains in pressure equilibrium with the surrounding plasma, and for this it must maintain a constant energy density (pressure), $U_{\rm{TOT}}$.

We assume an initial radius of $R_0 = 10^{10}~\rm{m}$. This value is $ \approx 10^4$ times smaller than typical radii used to model the whole jet for multiple blazar sources \citep[e.g.][]{2002A&A...386..833G,2013ApJ...768...54B} and our tests have shown individual plasmoids smaller than this size do not radiate enough to be observable and so will not be astrophysically relevant, justifying this assumption. We assume that the flaring emission is dominated by this single plasmoid, which is approximately spherical in its rest frame. To test whether reconnection is a feasible energy source for powering flares we allow the plasmoid to grow in the reconnection layer, assuming a growth rate which depends on the merge timescale, $\tau_{\rm{merge}}$, and relative blob velocity in the jet rest frame, $v_{\rm{b}}$,

\begin{equation}
  \tau_{\rm{acc}}=\tau_{\rm{merge}} = f\frac{R(t)}{v_{\rm{b}}(t)},
  \label{eq:tm}
 \end{equation}
where $f$ is a dimensionless free parameter quantifying the plasmoid spacing and relative velocity in the reconnection layer and $R$ is the radius. Following \citet{2016MNRAS.462...48S}, we obtain the blob velocity, $v_{\rm{b}} = \beta_{\rm{b}}c$, by solving, 

\begin{equation}
\Gamma_{\rm{b}} \beta_{\rm{b}} = \sqrt{\sigma_0} \tanh \left( \frac{0.12 x}{\sqrt{\sigma_0}R } \right),
\label{blobvel}
\end{equation}
where $x$ is the distance travelled by the blob in the reconnection layer and $\Gamma_{\rm{b}}$ is the Lorentz factor of the blob. We assume an initial value of $x$ in Eqn. \ref{blobvel} equal to the initial plasmoid radius. Rearranging this equation leads to, 

\begin{equation}
\beta_{\rm{b}} = \frac{\sqrt{\sigma_0} \tanh \left( \frac{0.12 x}{\sqrt{\sigma_0}R } \right)}{\sqrt{1+\left(\sqrt{\sigma_0} \tanh \left( \frac{0.12 x}{\sqrt{\sigma_0}R } \right) \right)^2}}
\label{eq:vblob}
\end{equation}
We assume that the plasmoid radius grows steadily whilst in the reconnection layer, see Fig. \ref{modelFig}. The rate of radial growth can be calculated from the condition that the plasmoid volume doubles in a time $\tau_{\rm{merge}}$, and is given by, 

\begin{equation}
\frac{{\rm d}R}{{\rm d}t} = \frac{2^{1/3}R - R}{\tau_{\rm{merge}}} =
\left(2^{\frac{1}{3}} - 1 \right) \frac{v_{\rm{b}}(t)}{f} \approx 0.26 \frac{v_{\rm{b}}(t)}{f}.
\label{eq:dRdt}
\end{equation} 
We assume $f$ cannot be less than one else single plasmoids will be unable to form. This gives a maximum possible radial growth rate of $0.26 c$. We assume that the value of $U_{\rm{e}}$ is constant during the merging process, such that electrons are injected over time at the same rate as the radial growth in Eqn. \ref{eq:dRdt}. This is the case if the two merging plasmoids have indentical properties. In the event of a negligible guide field where the emitting plasmoid becomes particle dominated with $\sigma<10^{-3}$, we assume that the emitting blob is no longer in the reconnection layer to define the final radius, $R_{\rm{f}}$, and total time, $t_{\rm{f}}$ spent in reconnection later in the plasmoid rest frame. We estimate a lower limit to the size of the reconnection layer, $L$, by computing the total distance travelled by the emitting plasmoid in the jet rest frame,

\begin{equation}
L = \int_t^{t_{\rm{f}}} v_{\rm{b}}(t) {\rm d}t,
\label{eq:L}
\end{equation}
where $t$ is the time as measured in the jet rest frame. 

\subsection{Particle Acceleration}

The objective of this work is to model emission from an accelerating and growing plasmoid in a reconnection layer. To accurately model radiative losses, the underlying electron population in the reconnecting plasmoid needs to be tracked as it undergoes particle acceleration. To restrict the number of free parameters, the initial electron population residing in the emitting plasmoid is approximated as a $\delta$-function with Lorentz factor $\gamma = 10$. We investigated different functional forms for the initial electron population, but found they had a negligible influence on the end result. From Eqn. \ref{eq:Ubal}, the initial magnetisation, $\sigma_0$, is used to set the initial value of $U_{\rm{e}}$ and therefore the electron number density in the plasmoid. 

\begin{figure}
	\includegraphics[width=\columnwidth]{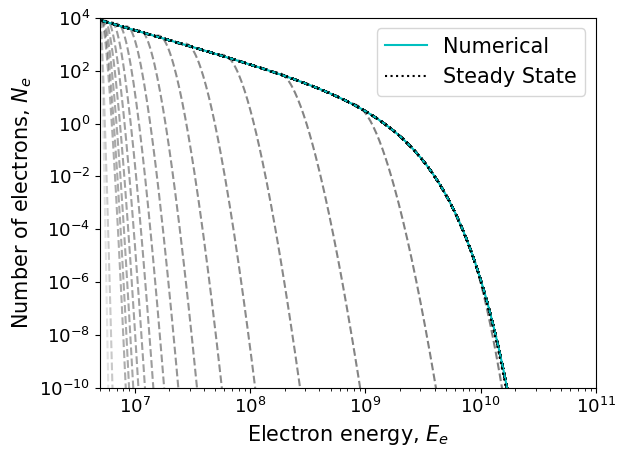}
    \caption{Demonstration of accuracy for the particle acceleration code. The electron population at various times is represented by the grey dashed line, which converges to the numerical solution indicated by the cyan line. The analytical steady state solution is indicated by the black dotted line, with only slight discrepancies to the numerical solution at low electron counts. The converged electron spectral index from our code differs with the analytical solution index by less than half a percent.}
    \label{particleEv}
\end{figure}

It is necessary to model the evolution of the electron spectrum to obtain the dynamically evolving radiative losses from them. In general, the resultant spectrum from a particle acceleration process can be found by solving the diffusion-loss equation \citep{Longair:2011},

 \begin{equation}
 \frac{\mathrm{d}N(E)}{\mathrm{d}t} = D\nabla^2N + \frac{\partial}{\partial E} \left[b(E)N(E) \right] - \frac{N(E)}{\tau_{\rm{esc}}} + Q(E),
 \label{DL_orig}
 \end{equation}
where the terms on the right hand side quantify diffusion, energy gains or losses, escape from the acceleration region and repeated particle injection. Charged particles in a uniform magnetic field undergo helical motion, with gyro-radius $r_g$ and period $\tau_g$ defined as \citep{Longair:2011},
 \noindent\begin{minipage}{0.2\textwidth}
 \begin{equation}
 r_{\rm{g}} = \frac{\gamma m_{\rm{e}}\left| v \right| }{e \left| B \right|} \sin \theta,
 \label{rg}
 \end{equation}
 \end{minipage}%
\begin{minipage}{0.08\textwidth}\centering
\hfill  
\end{minipage}%
 \begin{minipage}{0.2\textwidth}
  \begin{equation}
 \tau_{\rm{g}} = \frac{2\pi r_{\rm{g}}}{c},
 \label{taug}
 \end{equation}
  \end{minipage}%
  \vskip1em
\noindent where $v$ is the velocity of an electron of energy $E$, $B$ is the magnetic field strength and $\theta$ is the pitch angle between $B$ and the direction of motion. We assume a tangled field so that diffusion out of the region is approximately a random walk with mean free path $\approx r_{\rm{g}}$ (Bohm diffusion \citep[e.g.][]{1960PhFl....3..659S}). Since the plasmoid is travelling relativistically towards the observer and its emission is Doppler boosted, we ignore the un-beamed emission from electrons which have left the plasmoid as a first approximation. It is unknown how efficient magnetic reconnection is as an acceleration mechanism, so we parameterise our acceleration term as $b = - \alpha E/\tau_{\rm{merge}}$ \citep{Longair:2011}, where $\alpha$ is a dimensionless free parameter quantifying the average energy gain per particle per merging event, thus $b$ is the average energy gain per unit time in the model. Without particle injection, Eqn. \ref{DL_orig} becomes,
  
 \begin{equation}
 \frac{\mathrm{d}N(E)}{\mathrm{d}t} = - \frac{N(E)}{\tau_{\rm{g}}}\left(\frac{r_{\rm{g}}}{R}\right)^2 -\frac{\alpha N(E)}{\tau_{\rm{merge}}} - \frac{\alpha E}{\tau_{\rm{merge}}}\frac{\mathrm{d}N(E)}{\mathrm{d}E}  - \frac{N(E)}{\tau_{\rm{esc}}} + \dot{N}_{\rm{rad}},
 \label{DLeqn}
 \end{equation}
where the first term describes Bohm diffusion, where it is assumed that if the electron gyroradius exceeds the radius of the emitting plasmoid containing it, it can leave the plasmoid and no longer contributes to the emission. The second and third terms describe the electron energy gains due to acceleration, and conservation of particle number. The fourth term quantifies electrons which leave the acceleration region per unit time, and the final term describes population changes due to radiative losses, which are discussed in Section \ref{sec:radloss}. Ignoring the contribution of radiative losses, Eqn. \ref{DLeqn} can be solved analytically giving the standard steady state ($\dot{N}=0$) solution,

 \begin{equation}
N(E) = A E^{-\left(1+\frac{\tau_{\rm{merge}}}{\alpha \tau_{\rm{esc}}}\right)} \exp  \left(-\frac{\tau_{\rm{merge}}}{\alpha \tau_{\rm{g}}}\frac{r_{\rm{g}}^2}{R^2}\right) 
 \label{SSDL}
 \end{equation}
which we use as a simple test of numerical convergence for our code, demonstrated in Fig. \ref{particleEv}. Substituting in Eqn. \ref{eq:tm} and rewriting Eqn. \ref{rg} in terms of energy allows the value of the exponential cutoff, $E_{\rm{max}}$, to be determined as,

\begin{equation}
E_{\rm{max}} = E \left( \frac{\tau_{\rm{merge}}}{\alpha \tau_{\rm{g}}} \frac{r_{\rm{g}}^2}{R^2} \right)^{-1} = \frac{2 \pi e \alpha R B v_{\rm{b}}}{f},
\end{equation}
giving the intuitive result that larger blobs can hold higher energy electrons with larger gyro-radii. We do not expect a substantial number of electrons to escape from the reconnecting plasmoid as opposed to the case of a shock front, so we set $\tau_{\rm{esc}} = \infty$ \citep[see][]{2014PhRvL.113o5005G}. From Eqn. \ref{SSDL} the power law index on the electron spectrum should therefore converge $\approx 1$. The parameters $\alpha$ and $f$ are left as free.

\subsection{Radiative Losses}\label{sec:radloss}
\begin{figure}
\subfloat[Case 1: The plasmoid grows continually during its time in the reconnection layer. These plasmoids can grow large enough that their emitted radiation can be observed. ]{\includegraphics[width=0.45\textwidth]{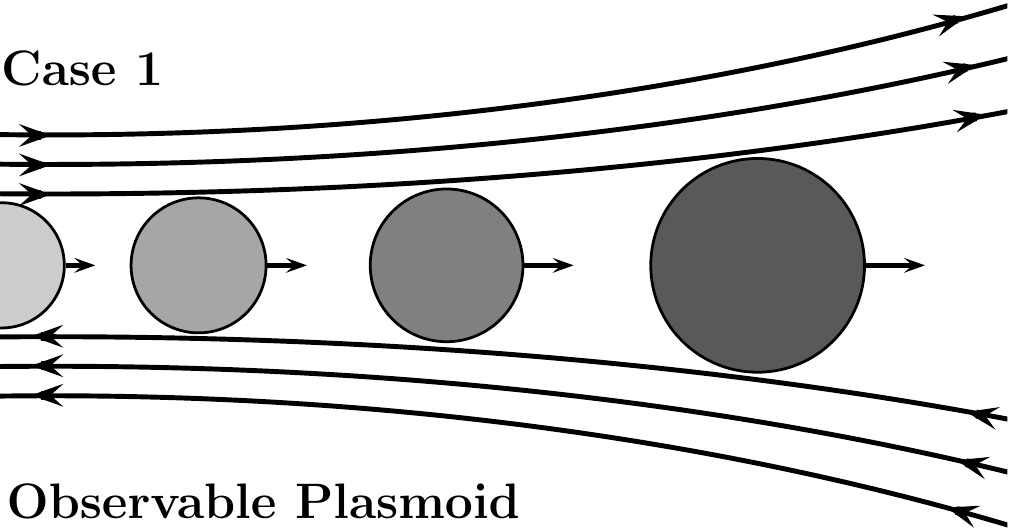}} \label{Case1} \\
\subfloat[Case 2: the radiative losses from the emitting electrons are so extreme that they cannot stay at high energies. The blob therefore shrinks continuously to maintain pressure equilibrium, and never grows large enough to be observed.]{\includegraphics[width=0.45\textwidth]{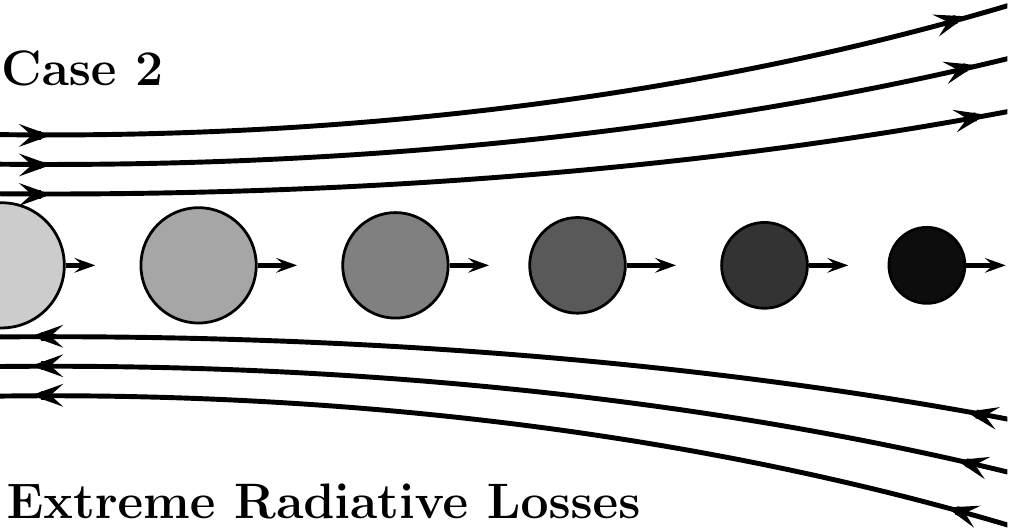}} \label{Case2} \\
\subfloat[Case 3: The plasmoid initially contains electrons with extreme radiative losses, causing the radius to contract to maintain pressure equilibrium with the surrounding plasma. From Eqn. \ref{eq:tm}, this reduces $\tau_{\rm{merge}}$, increasing the acceleration efficiency which supports the population of high energy electrons and allowing for some net growth. Since $\tau_{\rm{merge}}$ is the characteristic acceleration time, a smaller value also causes more rapid particle acceleration, which increases the total radiative losses. If the losses are extreme enough, this can cause a net contraction of the plasmoid radius. The radius of the emitting plasmoid oscillates to maintain pressure equilibrium with the surrounding plasma depending on whether the plasmoid merging or radiative losses are dominant. These plasmoids do not reach the required radii needed to be observable. This process does not happen indefinitely, as the plasmoid still suffers a loss of energy over time, which reduces the magnetic field and therefore the radiative losses.]{\includegraphics[width=0.45\textwidth]{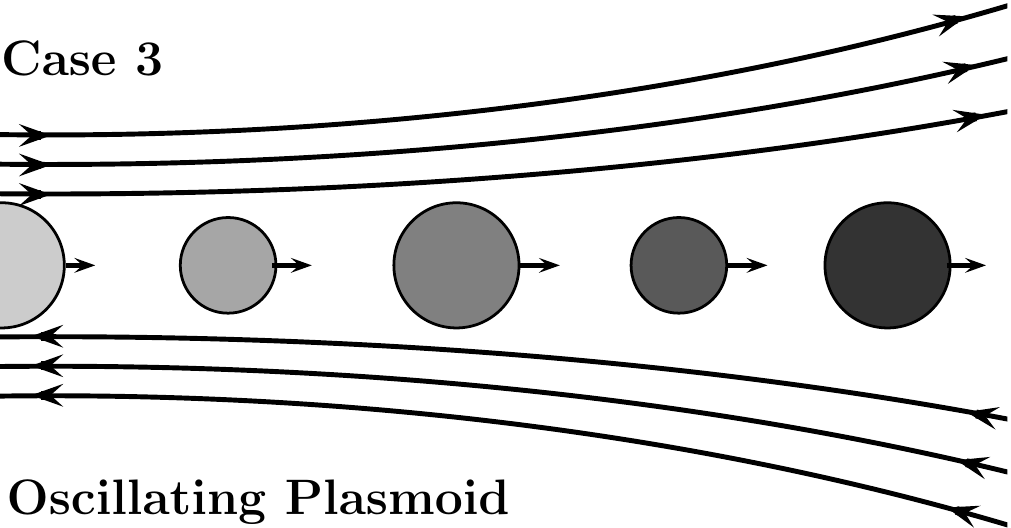}} \label{Case3} \\
\caption{The model presented here produces emitting plasmoids whose evolution typically follows one of the above scenarios.}
\label{fig:cases}
\end{figure}

We explicitly compute the emission and radiative losses from synchrotron and SSC processes, which we assume to be the dominant emission mechanisms from the reconnecting plasmoid. Our method follows that prescribed in \mbox{\citet{PC:2012}}. The synchrotron power for a radiating electron is given by,

\begin{equation}
P_{\rm{sync}} = \frac{4 \sigma_{\rm{T}}}{3 m_{\rm{e}}^2 c^3} \beta^2 E^2 U_{\rm{B}},
\label{Psync}
\end{equation}
where $\sigma_{\rm{T}}$ is the Thomson cross section. We assume that the synchrotron emission becomes optically thick when the optical depth is one, and use the Rayleigh -Jeans approximation for the synchrotron emission \citep{Longair:2011}. 
It is assumed that all electrons within a bin emit all of their energy at the critical frequency, defined as \citep[e.g.][]{Longair:2011},

\begin{equation}
\nu_{\rm{c}} = \frac{3 \gamma^2 e B}{4 \pi m_{\rm{e}}},
\label{eq:fcrit}
\end{equation}
and is the peak emission frequency of radiation emitted by an electron of Lorentz factor $\gamma$. Similarly, the inverse-Compton power is,
\begin{equation}
P_{\rm{IC}} = \frac{4 \sigma_{\rm{T}}}{3 m_{\rm{e}}^2 c^3} \beta^2 E^2 U_{\rm{rad}},
\label{P_IC}
\end{equation}
where we assume $U_{\rm{rad}}$ is the photon energy density of the synchrotron field, which is defined as

\begin{equation}
U_{\rm{rad}} = \left( 4 \pi R^2 l_{\rm{c}} \right)^{-1} \int_{E_{\rm{min}}}^{\infty} N_{\gamma}(E) {\rm d} E,
\label{eq:urad}
\end{equation}
where $N_{\gamma}(E)$ is the number of synchrotron seed photons with energy $E$. It is assumed the radiation emitted in one second is contained within a spherical shell of thickness one light second, $l_{\rm{c}}$, and volume $4 \pi R^2 l_{\rm{c}}$ \citep{PC:2012}. This leads to the definition of $\dot{N}_{\rm{rad}}$ in Eqn. \ref{DLeqn} as,

\begin{equation}
\dot{N}_{\rm{rad}} = \frac{{\rm d} \left[ P(E) N(E) \right] }{ {\rm d} E},
\end{equation}
which describes radiative losses for electrons of energy $E$. To solve this numerically, our electron population is divided into bins. Each bin gains electrons as the higher energy electrons radiate and reduce their energies and loses electrons which radiate energy. As we are particularly interested in modelling TeV flares, we need to account for Klein-Nishina effects which suppress the emission of high energy photons. The Klein-Nishina cross section is given by \citep[e.g][]{1970RvMP...42..237B},
\begin{equation}
\frac{\rm{d}\sigma}{\rm{d}\Omega} = \frac{1}{2} \alpha^2 r_{\rm{c}}^2 P^2 \left( P + P^{-1} - 1 + \cos^2 \phi \right),
\end{equation}
where $\alpha$ is the fine structure constant and $r_{\rm{c}}$ is the classical electron radius. The term $P$ is the energy ratio of the outgoing to incoming photon in the rest frame of the electron and is defined as,
\begin{equation}
P = \frac{1}{1+\frac{E_{\gamma}}{m_{\rm{e}}c^2}\left( 1 - \cos \phi \right)},
\end{equation}
where $\phi$ is the angle between the incident photon of energy $E_{\gamma}$ and the outgoing photon. After extracting the radiative energy losses from the emitting plasmoid, its radius will decrease to maintain pressure equilibrium with its surroundings. This effect leads to a rise in the electron number density in the plasmoid, and a loss of total energy over time. It ensures total energy conservation in the model and allows the emitting plasmoid to either shrink or grow, depending on the competing effects of radiative losses and plasmoid growth from merging. This leads to a number of possible scenarios, which are illustrated in Fig. \ref{fig:cases}.

\subsection{TeV Opacity}

Simulations of TeV flares are further complicated by the requirement that the TeV photons must be able to leave the emitting region, with regions containing dense photon fields being opaque to high energy photons \citep{1966PhRvL..16..252G}. It is possible for a high energy photon of energy $E_{\gamma}$ to interact with a lower energy photon to produce an electron-positron pair, with the threshold energy of the lower energy photon given by \citep{1967PhRv..155.1404G},
\begin{equation}
E_{\rm{thresh}} = \frac{2m_{\rm{e}}^2c^4}{E_{\gamma}(1-\cos\theta)},
\label{TeVthresh}
\end{equation}
where $\theta$ is the angle of incidence between the two photons. The cross section for $\gamma \gamma \rightarrow e^+ e^-$ is, 

\begin{equation}
\sigma_{\gamma \gamma} = \frac{1}{2}\pi r_{\rm{c}}^2\left(1-\beta^2\right) \left[ \left(3-\beta^4 \right) \ln \left( \frac{1+\beta}{1-\beta} \right) -2\beta \left(2-\beta^2 \right) \right],
\label{opeqn}
\end{equation}
where $\beta$ gives the velocity of the pair-produced electron-positron pair in the centre-of-mass frame and $r_{\rm{c}}$ is the classical electron radius. The total optical depth is then computed by \citep[e.g.][]{2013MNRAS.436..304P},

\begin{equation}
\tau_{\gamma \gamma} \left( E_{\rm{SSC}} \right) = \sum_{E=E_{\rm{thresh}}}^{\infty} \sum_{\theta=0}^{\theta=\pi} R \left( 1 - \cos \theta \right) n_{\rm{sync}} \frac{\sin \theta}{2} \sigma_{\gamma \gamma} {\rm d}E \rm{d}\theta,
\end{equation}
where $n_{\rm{sync}}$ is the photon number density from the synchrotron emission. The SSC emission is then attenuated by a factor of $\exp \left( -\tau_{\gamma \gamma}\left( E_{\rm{SSC}} \right) \right) $. Here we neglect the contribution of the electron pair cascade produced by this process because we find the observable plasmoids are largely optically thin to TeV radiation and because the TeV luminosity is much lower than the GeV luminosity, so only a relatively small number of pairs will be produced when compared with the existing electron population. 

\subsection{Doppler Boosting}

An important property of reconnection generated plasmoids is that they can have bulk Lorentz factors that exceed that of the jet, and thus emission from them can be more strongly Doppler boosted than the surrounding quiescent plasma. To calculate the observed Lorentz factor of the plasmoid, $\Gamma_{\rm{p}}$, we need to consider the relative components of its velocity aligned with and perpendicular to the jet motion. Following \citet{2016MNRAS.462.3325P}, the observed Doppler factor, $\Gamma_{\rm{p}}$, for a plasmoid travelling with Lorentz factor $\Gamma_{\rm{b}}$ at angle $\theta'$ relative to the jet axis in the rest frame of the jet is,

\begin{equation}
\Gamma_{\rm{p}} =  \Gamma_{\rm{j}} \Gamma_{\rm{b}} \left( 1 + \beta_{\rm{j}} \beta_{\rm{b}} \cos \theta' \right) = \left( 1 - \beta_{\rm{p}}^2 \right)^{\frac{1}{2}}.
\label{blobLor}
\end{equation}
The observed lab frame angle of the reconnecting plasmoid to the jet axis, $\theta$, is defined by,

\begin{equation}
\tan \theta = \frac{\beta_{\rm{b}} \sin \theta'}{\Gamma_{\rm{j}} \left( \beta_{\rm{j}} + \beta_{\rm{b}} \cos \theta' \right)},
\end{equation}
consequently giving a Doppler factor of,

\begin{equation}
\delta_{\rm{p}} = \frac{1}{\Gamma_{\rm{p}} \left( 1 - \beta_{\rm{p}} \cos \omega \right)},
\label{blobDopp}
\end{equation}
where the difference between the angle of the plasmoid to the observer and the observers angle to the jet axis is $\omega = \theta - \theta_{\rm{obs}}$. Using primes to denote the blob rest frame therefore results in amplification of $\nu_{\rm{obs}} F_{\nu, \rm{obs}} = \delta_p^{4+K} \nu_{\rm{emit}} F_{\nu, \rm{emit}}$, where $K$ is the k-correction.

\subsection{Code Setup}

The code utilises an explicit method with adaptive time-steps to solve the differential equation defined in Eqn. \ref{DLeqn}, which returns the accelerated electron spectrum. In the model, the constant evolution of the electron spectrum, dynamic magnetic field and growing radius of the emitting plasmoid mean that the system is not in equilibrium. As the emitting blob evolves, the time-step $\rm{d}t$ changes dramatically. This is because from Eqn. \ref{eq:tm} the characteristic merge time changes as the radius changes, and the shortest radiative lifetime is a function of the evolving energy spectrum (see Eqns. \ref{Psync} and \ref{P_IC}). Accordingly, using an adaptive time-step method was required to ensure the code ran fast and efficiently. This works by estimating the error in the discretised electron population by comparing the population after one time-step to that in two half time-steps. If the error is larger that one part in $10^6$, the time-step is halved to increase accuracy. If the error is below this tolerated value, the time-step was doubled to ensure speed and efficiency. This allowed the code to run in approximately 5 minutes. The code was written in C, and made use of the OpenMP module to parallelise the computations.

From Eqns. \ref{Psync} and \ref{P_IC}, the smallest radiative time-step is that of the highest energy electrons, which needs to be resolved by our code. To account for the computationally expensive nature of explicitly computing the radiative losses and Klein-Nishina cross section, the radiative losses are only recomputed if the highest energy bin changes, if the radius of the emitting plasmoid has changed by more than 5\% or if the minimum electron radiative lifetime is smaller than the acceleration time ${\rm d}t$ discussed previously to prevent electrons with fast radiative lifetimes emitting more energy then they have. The code utilises two sets of electron bins. A fine logarithmic grid of adjacent bin ratios 1.03 is used to accurately compute the particle acceleration, and to save computation time these were interpolated to a coarse grid each containing exactly 9 fine bins to compute the radiative losses. These relative widths were chosen as they provided significant speed up to the code and returned results indistinguishable to those relative to using finer bins. Computation of radiative losses is done with discrete electron bins and employs a similar methodology to \mbox{\citet{PC:2012}}.

\section{Results and Discussion}\label{sec:resdisc}

\subsection{Parameter Search Results}

To investigate the range of resultant SEDs that our reconnection plasmoids could produce, an investigation over the entire parameter space was performed, outputting approximately 10,000 SEDs. This parameter sweep included magnetic field values from $10^{-7} - 10^{-3}~\rm{T}$ and initial magnetisation values in the range $10 - 10^{6}$, each of which were logarithmically spaced in the parameter sweep. We investigated acceleration parameters in the range $\alpha = 1-20$ and $f=1-20$. Both of these were linearly spaced. From the resultant SEDs, we note the following general properties: 

\begin{itemize}
\item All of the produced SEDs had emission that was synchrotron-dominated.
\item The model can produce different characteristic flare profiles. These either have a slow rise and rapid decline, or vice versa (see Fig. \ref{fig:BF} and Fig. \ref{magflare}). We find that that the profile of the flare depends on $\tau_{\rm{merge}}$ and the magnetisation. This is discussed in Section \ref{sec:flareProf}.
\item The maximum electron energy is determined by an equilibrium between particle acceleration and radiative losses.
\item The power in SSC emission follows the relation $P_{\rm{SSC}} \propto R$ (see Eqns. \ref{eq:uradproptoR} and \ref{eq:PproptoR}) with a full discussion in Section \ref{sec:syncdom}. It follows that larger plasmoids output proportionally more SSC than plasmoids with smaller radii, and produce SEDs with increased Compton-dominance. Subsequently, plasmoids that produce enough SSC to be observable in $\gamma$-rays need to be large.
\item For plasmoids large enough to produce enough SSC to be observable, Eqn. \ref{eq:vblob} leads to blob velocity profiles that peak in the mildly relativistic regime. Plasmoids were found to initially accelerate from their birth locations, eventually reaching a peak velocity. It was possible for them to slow down slightly at late times.
\item Smaller, relativistic plasmoids could be produced but emission from them was heavily synchrotron dominated so they were unable to provide substantial radiative emission in the TeV regime.
\end{itemize}
We discuss each of these in the following sections.

\subsection{Dependence on Free Parameters}\label{disc:freeparam}

\begin{figure*}
        \centering
	\includegraphics[width=17.5cm]{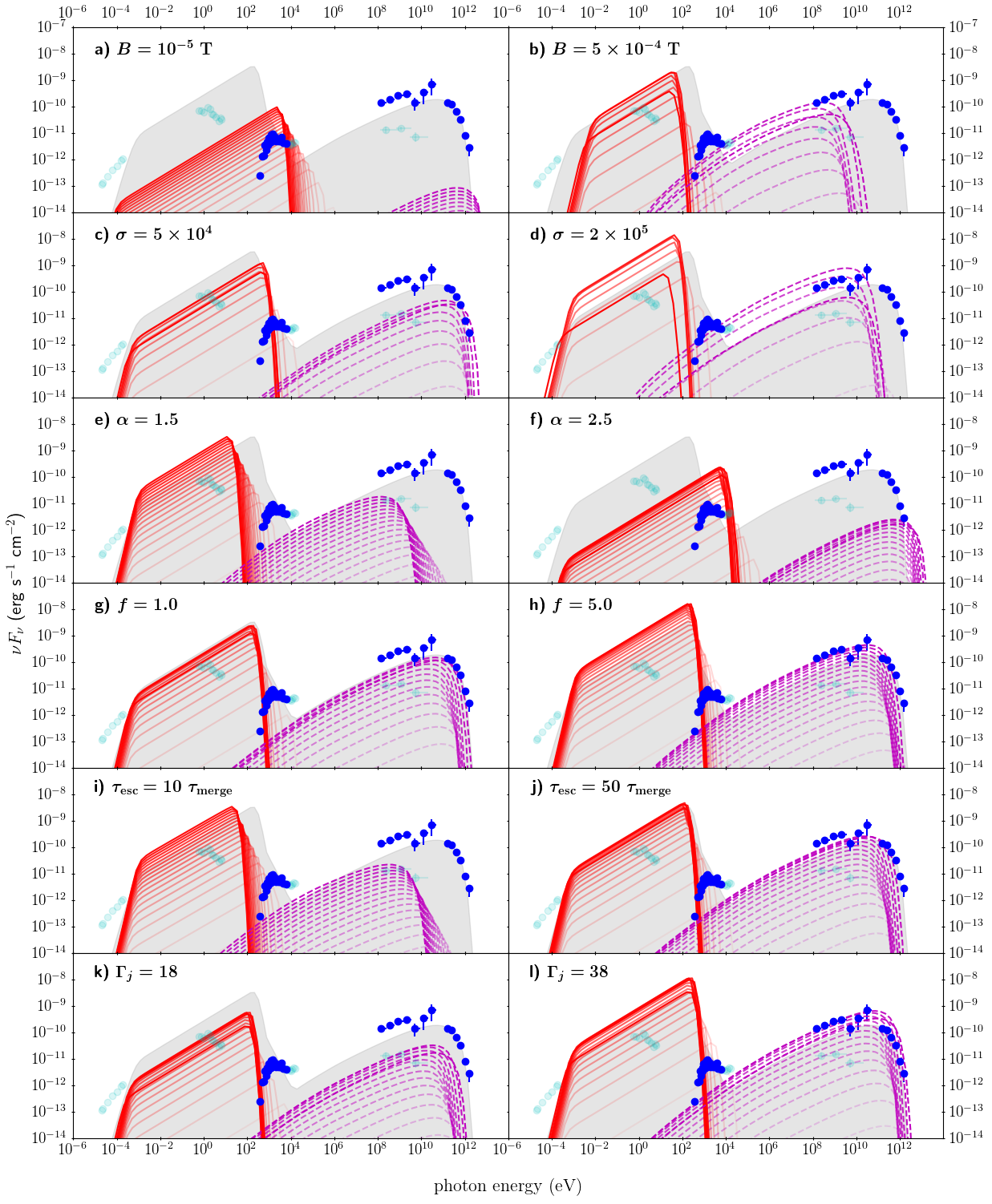}
    \caption{Figure showing how the time-dependent SEDs vary with respect to each model parameter. All changes are relative to the fit in Fig. \ref{fig:BF}, of which best fit parameters are listed in Table \ref{tab:BF}. This fit is shown by the grey silhouette in each panel to allow comparison of the evolution of the SEDs with respect to each changed parameter. Only one parameter was changed per panel, and is specified in each. The final radii and plasmoid rest frame times are defined when the magnetic energy in the reconnecting field has been completely depleted and are given in Table \ref{comptable}. Fainter colours represent SEDs produced at earlier times. The red solid lines are synchrotron radiation, and the dashed magenta lines SSC. The effects of changing these parameters are discussed in Section \ref{disc:freeparam}. The flaring data is taken from \citet{2018ApJ...856...95A} and is indicated by the dark blue circles. Quiescent data is show by the faint cyan circles for comparison \citep{2011ApJ...730..101A}.}
    \label{paramcomp}
\end{figure*}

\begin{table} 
	\centering
	\caption{Table listing the final radius and time the plasmoid was in the reconnection layer as measured in its rest frame, for the fits displayed in Fig. \ref{paramcomp}a-j.} 
	\label{comptable}
	\begin{tabular}{lcc} 
		\hline
		Figure & Final Radius $R_{\rm{f}}$ (m) & Duration (days) \\
		\hline
		\ref{fig:BF} &  $1.2 \times 10^{14}$ & 46.3 \\
		\ref{paramcomp}a & $1.4 \times 10^{14}$ & 56.6 \\
		\ref{paramcomp}b & $7.0 \times 10^{12}$ & 2.9 \\
		\ref{paramcomp}c & $7.3 \times 10^{13}$ & 28.7 \\
		\ref{paramcomp}d & $2.2 \times 10^{14}$ & 86.8 \\
		\ref{paramcomp}e & $1.4 \times 10^{14}$ & 56.7 \\
		\ref{paramcomp}f & $2.8 \times 10^{13}$ & 11.1 \\
		\ref{paramcomp}g & $1.3 \times 10^{14}$ & 43.8 \\
		\ref{paramcomp}h & $6.8 \times 10^{13}$ & 56.7 \\
		\ref{paramcomp}i & $1.4 \times 10^{14}$ & 56.7 \\
		\ref{paramcomp}j & $1.4 \times 10^{14}$ & 53.7 \\

		\hline
	\end{tabular}
\end{table}

Fig. \ref{paramcomp} and Table \ref{comptable} demonstrate how the observed radiation from the reconnection model varies with respect to changing each of the five main parameters. The maximum electron energy was determined from a balance between particle acceleration and radiative losses. This dynamically determines the maximum synchrotron and SSC emitted photon energies, which can be seen to vary through Fig. \ref{paramcomp}a-f. It can be seen that small changes in each parameter can significantly alter the resultant SED. We begin by summarising and explaining the behaviour of our code with respect to each of them.

\begin{figure}
	\includegraphics[width=\columnwidth]{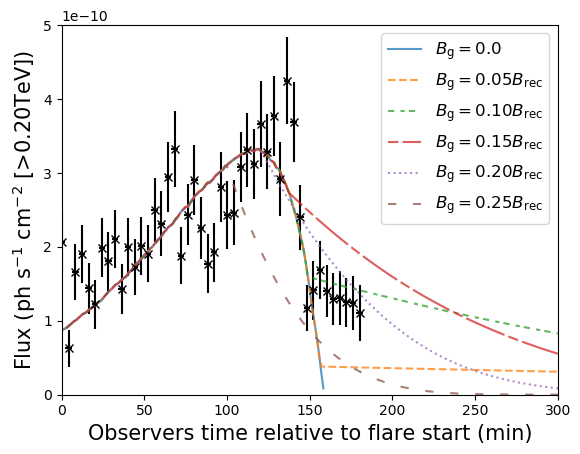}
    \caption{Figure displaying the effects of varying the relative strength of the magnetic guide field on the best fit light curve. The total initial magnetic field, $B = B_{\rm{rec}} + B_{\rm{g}}$, is constant for each simulation, with the relative strength of the guide field adjusted. As energy is transferred from the reconnecting field to particles, $B_{\rm{rec}}$ is depleted so the total field strength approaches that of the guide field. When $B_{\rm{g}}$ dominates, the decay of the flare is governed by the shortest radiative lifetime, which is shorter for stronger guide fields. Of these, the fit with $B_{\rm{g}}=0.1B_{\rm{rec}}$ gives the best fit, with $\chi_{\nu}^2 = 1.97$.}
    \label{fig:Bgdecay}
\end{figure}

\begin{itemize}
\item \textbf{Initial magnetic field,} $\boldsymbol{B_{\rm{0}} =} \boldsymbol{B_{\rm{rec,0}} + B_{\rm{g,0}}}$: 
Fig. \ref{paramcomp} shows that an initially higher magnetic field lowers the maximum emitted energy from both synchrotron and SSC, with the synchrotron emission becoming optically thick at a higher frequency. From Eqn. \ref{Psync}, the $B$-field dependancy of the synchrotron power is $P_{\rm{sync}} \propto E^2 B^2$, thus the radiative losses are higher for all electron energies and this effect is more profound for the highest electron energy populations. Therefore, if $B_{\rm{0}}$ is increased for constant acceleration parameters, the balance between particle acceleration and radiative losses necessarily occurs at a lower electron energy, reflected by the lower maximum energy in the SED curves in Fig. \ref{paramcomp}b. This property also leads to a denser synchrotron photon field, and therefore a correspondingly higher SSC peak, as $P_{\rm{SSC}} \propto U_{\rm{rad}}$. Table \ref{comptable} shows that the both the values of the final radius, $R_{\rm{f}}$, rest frame run time $t_{\rm{f}}$ and reconnection layer size, $L$, for Fig.  \ref{paramcomp}b are much lower than for Fig.  \ref{paramcomp}a. This arises because the larger radiative losses associated with a higher $B_0$ at the same $\sigma_0$ mean that the total energy is more quickly depleted leading to a smaller final radius and a faster time to reach $\sigma=0$. Additionally, the larger radiative losses associated with a larger $B$ cause more rapid blob shrinking, restricting the net growth of the emitting plasmoid. The location of the optically thin to thick boundary depends on the opacity, $k_{\rm{\nu}} \propto j_{\rm{\nu}}\nu^{-\frac{5}{2}}$, and path length $R$.  Since the emissivity $j_{\rm{\nu}}$ increases with $B^2$, the smaller radius for larger magnetic fields is not enough to offset the increased $k_{\rm{\nu}}$, thus the emitting blob becomes optically thick at higher frequencies.

The relative contributions of $B_{\rm{g}}$ and $B_{\rm{rec}}$ to the total magnetic field $B$ serve govern the late stage evolution of radiative emission from the plasmoids. Once the energy available for particle acceleration initially stored in $B_{\rm{rec}}$ has been transferred to the electrons, particle acceleration no longer occurs and it is assumed that the plasmoid is no longer residing within the reconnection layer. In this event, the plasmoid radiates with a constant magnetic field equal to $B_{\rm{g}}$, and the electron population evolves such that higher energy electrons which have shorter radiative lifetimes radiate their energy more quickly than electrons at lower energies. The effect which the relative contribution of the guide field to the total magnetic field has on the light curve is shown in Fig. \ref{fig:Bgdecay}.

\item \textbf{Initial Magnetisation,} $\boldsymbol{\sigma_0}$:
Figs \ref{paramcomp}c and \ref{paramcomp}d show a higher maximum emitted energy in the synchrotron and SSC peaks for a lower $\sigma_0$. Whilst at first this may seem counterintuitive, a lower magnetisation provides a reduced energy budget for particle acceleration, and this supply of energy is depleted more quickly, causing a more rapid reduction of the magnetic field than would be the case for higher $\sigma_0$. As $P_{\rm{sync}} \propto E^2 B^2$, the balance between radiative losses and particle acceleration now occurs at higher electron energies for lower $\sigma_0$. Table \ref{comptable} shows Fig. \ref{paramcomp}c has a smaller $R_{\rm{f}}$ and faster $t_{\rm{f}}$ relative to Fig. \ref{paramcomp}d, which is a consequence of a smaller total energy supply which is more rapidly exhausted. Thus extremely low $\sigma$ plasmoids are unlikely to grow large enough to be detectable. 

\item \textbf{Acceleration parameter,} $\boldsymbol{\alpha}$:

The parameter $\alpha$ defines the efficiency of the acceleration process. Figs \ref{paramcomp}e and \ref{paramcomp}f show that even a relatively small change in $\alpha$ substantially alters the resultant SED. The higher value of  $\alpha$ causes a more rapid transfer of magnetic to particle energy, more rapidly reducing the magnitude of the magnetic field, much like in Fig. \ref{paramcomp}c. Thus, higher $\alpha$ leads to smaller plasmoids with a higher electron energy cutoff and a correspondingly higher maximum emission energy. Taking this result with the magnetic field dependence in Figs \ref{paramcomp}a and b, the plasmoid is therefore capable of reaching larger radii when the magnetic field is low and the acceleration is inefficient. Table \ref{comptable} shows a smaller value of $\alpha$ produces larger reconnection plasmoids, and comparing Figs \ref{paramcomp}e and \ref{paramcomp}f demonstrates larger plasmoids exhibit greater radiative emission. Thus, if the value of $\alpha$ is too high, although the electron spectrum extends to high energies, the final size of the emitting plasmoid is too small for it to radiate enough to be observable.

\item \textbf{Filling Factor,} $\boldsymbol{f}$:

An increase in $f$ physically corresponds to more space between adjacent plasmoids in the reconnection layer, and increases $\tau_{\rm{merge}}$ from Eqn. \ref{eq:tm}. For a larger $f$, we therefore expect to see a reduction in the maximum achievable energy as the acceleration takes longer and so can be balanced by weaker radiative losses, which is apparent from Fig. \ref{paramcomp}e. This figure also shows a decrease across all emission energies, which is caused by the lower final radius. Table \ref{comptable} shows that whilst the final radius is comparable to that of the original fit, it is achieved in a longer timescale, which is a direct consequence of larger $f$ increasing $\tau_{\rm{merge}}$. Eqn. \ref{eq:vblob} shows that $\beta_{\rm{p}}$ decreases as $R$ increases. Eqn. \ref{eq:dRdt} shows that a small $f$ increases the rate of radial growth, thus $f$ influences $v_{\rm{blob}}$ and $\delta_{\rm{p}}$. These effects will be discussed further in Section \ref{sec:PIC}.

\item \textbf{Escape Time,} $\boldsymbol{\tau_{\rm{esc}}}$ : 
Although the escape time was not a free parameter and was set to $\tau_{\rm{esc}} = \infty$ we investigated the effect of it on the resultant SEDs. To do this, we assumed $\tau_{\rm{esc}}$ was some factor multiplied by $\tau_{\rm{merge}}$, indicated in Figs \ref{paramcomp}i-j. Fig. \ref{paramcomp}i shows that significant particle escape occurs for $\tau_{\rm{esc}} = 10 \tau_{\rm{merge}}$, which significantly reduces the high energy SSC emission. We find that if $\tau_{\rm{esc}} < \tau_{\rm{merge}}$ the amount of particle escape is so severe that no acceleration occurs. Fig. \ref{paramcomp}j shows that for the case of $\tau_{\rm{esc}} \gtrapprox 50~\tau_{\rm{merge}}$ there is very little change to the $\tau_{\rm{esc}} = \infty$ best fit.

\item \textbf{Jet Bulk Lorentz Factor,} $\boldsymbol{\Gamma_{\rm{j}}}$:
The parameters $\Gamma_{\rm{j}}$, $\theta_{\rm{obs}}$ and $\theta'$ serve to influence $\delta_{p}$, defined in Eqn. \ref{blobDopp}. They therefore all affect the resultant SED similarly, thus we only show the effect of $\Gamma_{\rm{j}}$. Changes in $\delta_{p}$ consequently shift $\nu F_{\nu}$ by $(\delta'/\delta)^{4+k}$ and the emitted photon energies by $\delta'/\delta$, where $\delta'$ is the new Doppler factor.

\end{itemize}

\subsection{Fitting to the 2016 TeV Flare of BL Lacertae}

\begin{table} 
	\centering
	\caption{Table listing best fit parameters for Fig. \ref{fig:BF} and Fig. \ref{magflare}. The parameters for Fig. \ref{fig:BF} correspond to a total energy density of $U_{\rm{TOT}} = 10^{-4}$~J m$^{-3}$. The corresponding mean observed plasmoid Lorentz factor is $\Gamma_{\rm{p}}=37 $, with the mean Doppler factor $\delta_{\rm{p}}=73$.  }
	\label{tab:BF}
	\begin{tabular}{lccc} 
		\hline
		Parameter & Fig. \ref{fig:BF} & Fig. \ref{magflare} \\
		\hline
		$B_0$ (T) & $5\times 10^{-5}$ & $5\times 10^{-5}$\\
		$B_{\rm{g}}$ (T) & $4.5\times 10^{-6}$ &$1\times 10^{-6}$ \\
		$\sigma_0$ & $9\times 10^{4}$ & $9\times 10^{5}$  \\
		$\alpha$ & 2.0 & 2.0 \\
		$f$ & 1.6 & 1.6 \\
		$\Gamma_{\rm{j}}$ & 28 & 28 \\
		$\theta'$ & $87^{\circ}$ & $87^{\circ}$ \\
		$\theta_{\rm{obs}}$ & $1.2^{\circ}$ & $1.2^{\circ}$ \\
		\hline
	\end{tabular}
\end{table}

\begin{figure*}
\subfloat[Best fit SED to the 2016 TeV flare of BL Lacertae. Synchrotron emission is indicated by the red solid line, and SSC by the dashed magenta line. Darker shades symbolise the radiative emission from later times. The flaring SED is that from \citep{2018ApJ...856...95A}. The quiescent SED \citep{2011ApJ...730..101A} is also shown by faint cyan dots for reference. We find that our reconnecting plasmoid model is not able to produce sufficient SSC emission to fit well to both low and high energy data.]{\includegraphics[width=0.75\textwidth]{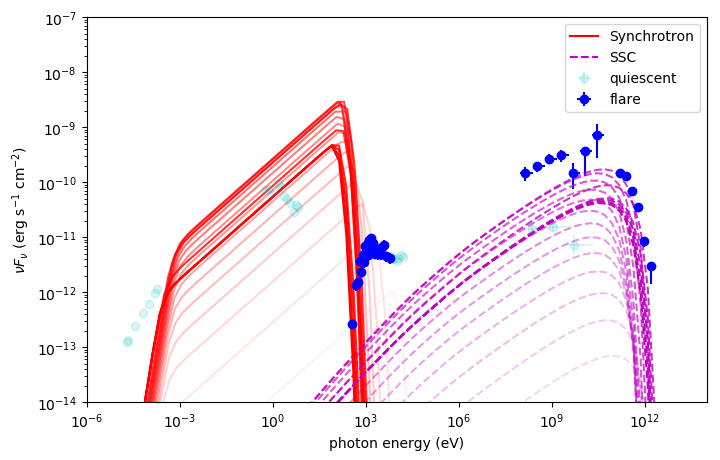}} \label{SEDfit} \\
\subfloat[Corresponding best fit to the VERITAS light curve of the 2016 BL Lacerate TeV flare \citep{2018ApJ...856...95A}. The fit has $\chi_{\nu}^2 = 1.95$, and is consistent with the SED in panel (a). It can be seen our model is able to reproduce the slow rise and rapid decline well. After $t\approx 145$~min, the energy stored in the reconnecting field has been entirely transferred to particles and the light curve evolves with $B=B_{\rm{g}}$.]{\includegraphics[width=0.47\textwidth]{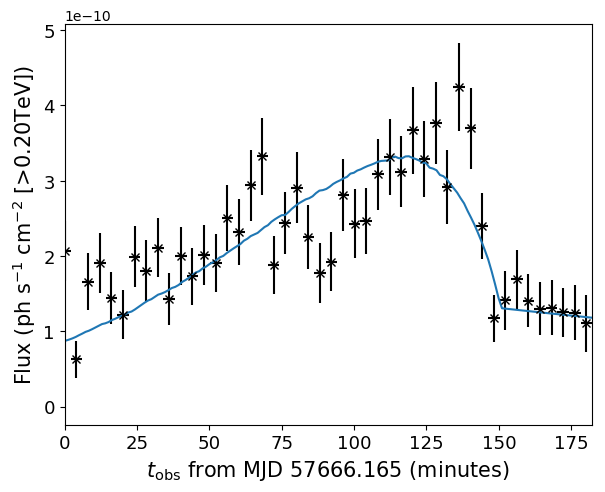}}\label{parflare} 
\hspace{0.5cm}
\subfloat[Figure showing the transfer of magnetic energy into particle and radiation energy for the above SED. It can be seen that the total energy density remains constant in the model, and that the radiation energy density $U_{\rm{rad}}$ never exceeds $U_B$, which is why the SEDs are synchrotron dominated. The grey region represents the rest frame time coincident with the flare shown in Fig. \ref{fig:BF}b. For $t \gtrapprox 900$~min, $B_{\rm{rec}}$ has been completely depleted and the value of $U_{\rm{B}}$ is the magnetic energy density stored in $B_{\rm{g}}$. In this time period, $U_{\rm{rad}}$ declines slightly as the plasmoid radius contracts as it emits radiatively.]{\includegraphics[width=0.48\textwidth]{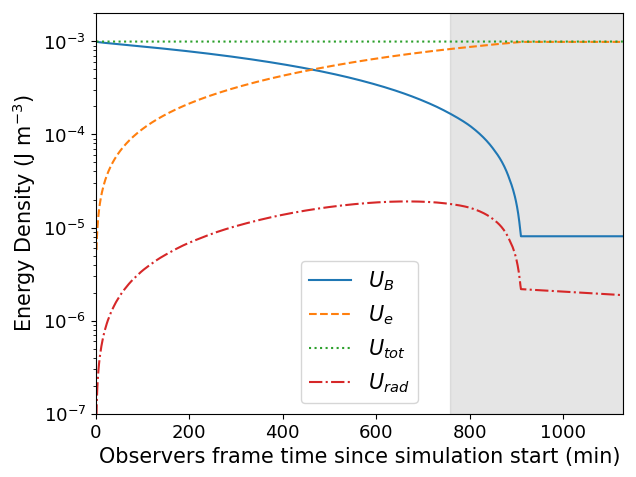}} \label{flareUtrack} \\
\caption{Simultaneous figs for the TeV light curve of BL Lacerate, the corresponding SED, and the tracking of the various energy densities in the system. The radial growth of the emitting plasmoid for this simulation proceeded approximately linearly, with any shrinking effects due to radiative losses negligible before $t \sim 900$~min when the reconnecting field has been depleted and the plasmoid no longer grows from merging as it is assumed to have left the reconnection layer. The plasmoid radial growth was approximately linear in time, as in Case A from Fig. \ref{fig:cases}.}
\label{fig:BF}
\end{figure*}


Magnetic reconnection has been postulated to account for the rapid variability associated at VHE frequencies. To assess the feasibility of reconnection as a mechanism which could power these flares, we applied our model to the 2016 TeV flare of BL Lacerate \citep{2018ApJ...856...95A}. We simultaneously fitted the SED and TeV light curve, which can be seen in Fig. \ref{fig:BF}. The free parameters of the best fit are given in Table \ref{tab:BF}. To obtain best fits to the data, we then selected the SEDs from our parameter search (10,000 runs) with SED fits closest to the BL Lacertae SED of \mbox{\citep{2018ApJ...856...95A}} and refined the fit by eye. Since the SED fit is relatively poor a $\chi^2_{\nu}$ minimisation is not very easy to apply given the lack of simultaneous data at different wavelengths.

When converting our luminosity values into observed fluxes, we adopt a redshift for BL Lacertae of $z=0.0686$ \citep{1995ApJ...452L...5V}, corresponding to a co-moving distance of $289~\rm{Mpc}$ assuming a flat $\Lambda$CDM Universe with $H_0 = 70~\rm{km s}^{-1}~\rm{Mpc}^{-1}$ \citep[e.g.][]{2013ApJS..208...20B}.

Fig. \ref{fig:BF}c indicates that the reconnection plasmoid required to fit this particular TeV flare is particle dominated with $\sigma<1$ during the time the majority of the TeV photons are emitted. We note that although the best fit to the 2016 TeV flare required a particle dominated plasmoid, flaring in our model does not uniquely occur when reconnecting plasmoids are in this regime. Flaring profiles corresponding to different dominant energy densities in the emitting plasmoids are discussed in Section \ref{sec:flareProf}. Fig. \ref{fig:BF}c shows that $U_{\rm{e}}$ for this plasmoid increases by over a factor of 100, which is necessary for the plasmoid to contain electrons which have undergone the substantial particle acceleration required for them to produce SSC emission at TeV energies. The plasmoid reaches equipartition at around $t=450$~min, after which it becomes particle dominated as $B_{\rm{rec}}$ continues to be depleted as particles are accelerated. As expected, when in equipartition, we find that the bolometric rest frame luminosity of the plasmoid is highest because the energy is evenly distributed between particles and magnetic fields. Beyond this, the bolometric luminosity decreases but the Compton-dominance of the SED increases as the reconnecting magnetic field weakens, reducing the radiative losses and increasing the maximum electron energy available for inverse-Compton scattering. We therefore find that SEDs are more strongly synchrotron dominated when $\sigma > 1$. We have also investigated the effect of having reconnecting plasmoids that remain in equipartition after the initial particle acceleration. However, we find that the conditions in the plasmoid required to best account for extreme transient TeV emission mean that either the reconnecting magnetic field is declining and reducing $U_{\rm{B}}$, or that the electrons are strongly radiating, which decreases $U_{\rm{e}}$. Whichever of these two effects is dominant, the emitting plasmoid quickly moves out of equipartition. Therefore, to maintain equipartition in our model we would have to inject $U_{\rm{B}}$ or $U_{\rm{e}}$, which would not conserve energy.


\begin{figure}
	\includegraphics[width=\columnwidth]{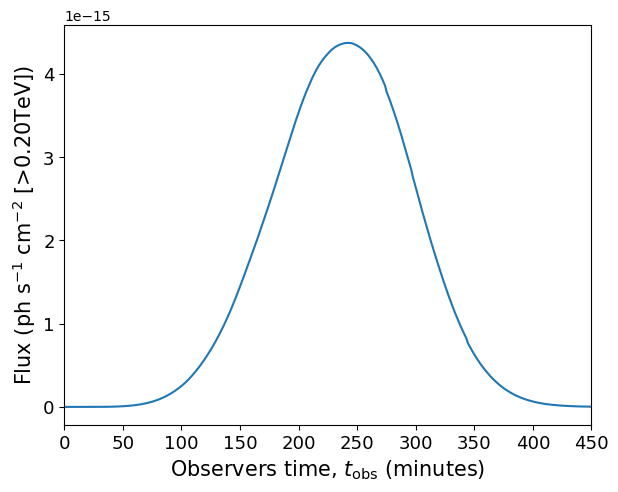}
    \caption{Typical flare profile for a magnetically dominated regime. It can be seen that the characteristic decay time of the flare is comparable to the rise time. The initial parameters are given in Table \ref{tab:BF}. It can be seen that changes in $\sigma_0$ and $B_{\rm{g}}$ are the only difference in parameters between this flare and the best fit flare in Fig. \ref{fig:BF}, yet they have very different profiles. It should be noted unlike for Fig. \ref{fig:BF} that the free parameters for Fig. \ref{magflare} have not been optimised in any way, and are included to provide an example of a flare originating from a magnetically dominated emitting plasmoid.}
    \label{magflare}
\end{figure}

\subsection{Synchrotron Dominated SEDs}\label{sec:syncdom}
It is clear from Fig. \ref{fig:BF}a that the best fit SED to the BL Lacertae 2016 flare is synchrotron dominated, with the X-ray synchrotron peak $\nu F_{\nu}$ value 10 times greater than the SSC peak. This synchrotron-dominance was a general property of all SEDs produced in our simulations, and our model is unable to produce reconnection plasmoids with Compton-dominated SSC emission. Whilst no simultaneous synchrotron data was recorded during the 2 hour TeV flare and therefore the synchrotron emission is unconstrained, our model predicts X-ray emission 3 orders of magnitude higher than that observed the day after the flare. This is over 10 times higher than typical X-ray emission observed from BL Lacertae \citep{2002babs.conf...63G}, thus it is likely that the model over-predicts synchrotron emission relative to observations. This is consistent with the model of \citet{NGBUS:2011}, which found that their radiative mini-jet model also over-predicted X-ray emission.

The model presented here converts magnetic energy into particle energy by accelerating the electron population residing in the reconnecting plasmoid at the expense of the energy in the magnetic field. The constituent electrons in the plasmoid then radiate via synchrotron and SSC. To obtain a Compton dominated SED, from Eqns \ref{Psync} and \ref{P_IC}, we require $U_{\rm{rad}} > U_{\rm{B}}$. Fig. \ref{fig:BF}c shows that this was not the case for the best fit model, and this result was true in all of our calculations. For each set of free parameters, whilst $\sigma > 1$, $U_{\rm{B}} > U_{\rm{e}} \geq U_{\rm{rad}}$, thus the peak radiative losses from synchrotron necessarily exceed that from SSC. Fig. \ref{fig:BF}c demonstrates that it is possible for $U_{\rm{e}} > U_{\rm{B}}$, but the decline in $U_{\rm{B}}$ leads to a fall in the strength of the magnetic field, $B$. As the magnetisation of the reconnecting plasmoid approaches zero, where it is completely particle dominated, the magnetic field decreases rapidly and substantially, as indicated in Fig. \ref{fig:BF}c after 800 minutes. This depletes the photon field provided from the synchrotron emission, and therefore $U_{\rm{rad}}$ also suffers the same rate of decline as $U_{\rm{B}}$, and thus can never exceed it so that our SEDs can never be Compton dominated. Furthermore, assuming no external photons are present, if $U_{\rm{rad}}/U_{\rm{B}} > 1$ then from Eqns. \ref{Psync} and \ref{P_IC}, the SSC losses are necessarily higher than those caused by synchrotron emission. Here, the majority of synchrotron seed photons, even those at low energies, become up-scattered to X-rays or $\gamma$-rays. The corresponding energy losses are therefore significant ($P \propto E^2$), thus the underlying electron population quickly loses energy and reverts to a lower energy configuration. In our model, these extreme losses would further cause the emitting plasmoid to shrink, decreasing the chances of it becoming large enough to be observable.

Between 600-800 minutes, Fig. \ref{fig:BF}c shows that $U_{\rm{rad}}$ declines less rapidly than $U_{\rm{B}}$. This is a consequence of the radial growth of the emitting plasmoid. Larger plasmoids contain more radiating electrons and so have larger $U_{\rm{rad}}$, and to some extent this can compensate for the decline in $B$. The synchrotron emissivity is proportional to the number of emitting electrons and this is proportional to the volume of the radiatively emitting plasmoid, i.e. $j_{\nu} \propto R^3$. In the calculation of the SSC emission, it is assumed that synchrotron seed photons are emitted into a spherical shell of thickness ${\rm d}R = l_{{\rm c}}$, where $l_{\rm{c}}$ is one light second \citep{PC:2012}, which is given in Eqn. \ref{eq:urad}. In Eqn. \ref{eq:urad}, $N_{\gamma} \propto N_{\rm{e}}$, thus the total radiation energy density provided by a fixed population of electrons producing synchrotron seed photons is given by,

\begin{equation}
  U_{\rm{rad}} \propto N_{\rm{e}}/V_{\rm{shell}} \propto R^3/R^2 \propto R.
  \label{eq:uradproptoR}
 \end{equation}
Combining Eqn. \ref{eq:uradproptoR} with Eqn. \ref{P_IC} yields, 
\begin{equation}
P_{\rm{IC}} \propto U_{\rm{rad}} \propto R. 
\label{eq:PproptoR}
\end{equation}
Hence the maximum Compton dominance we can achieve from our model is for large plasmoids approaching particle dominance. In PIC simulations, smaller plasmoids reach more relativistic speeds than larger ones \citep{2016MNRAS.462...48S}, thus radiative emission from them is more strongly Doppler boosted for an observer relative to larger plasmoids if both are beamed towards the observer. The above argument leading to Eqn. \ref{eq:PproptoR} regarding the ratio of SSC to synchrotron emission justifies choosing our own initial value of plasmoid radius to be relatively large because a large radius is a requirement to produce the amount of SSC $\gamma$-ray emission observed.

It is clear that adding in external photon fields and computing the contribution of external inverse-Compton (EIC) radiation may help alleviate the issue that the plasmoid emission is synchrotron dominated. In this work, we wished to establish initially whether the simplest possible case of SSC emission alone could account for rapid TeV flaring in blazars, in much the same way as it can explain the SEDs observed from many BL Lac type blazars \citep[e.g.][]{2013MNRAS.431.1840P}. The blazar flare fitted with our reconnection model in this paper is BL Lacertae. Although defines the BL Lac blazar subclass of blazar, it does not always behave as a BL Lac and has evidence for a broad line region (BLR) \citep{1995ApJ...452L...5V}. Free electrons in the BLR could potentially scatter external photons in the jet, providing additional seed photons to be inverse-Compton scattered. Another possibility is that other plasmoids, or emission from the surrounding current layer \citep{2013MNRAS.431..355G} provide additional photons for EIC. The addition of EIC would likely increase the Compton-dominance of emitting plasmoids in our reconnection model, which in turn may make the contribution to the emission from smaller plasmoids important. Investigating these effects therefore represents a significant amount of work which was considered beyond the scope of this paper, but is something we aim to explore in future work.

\begin{figure}
	\includegraphics[width=\columnwidth]{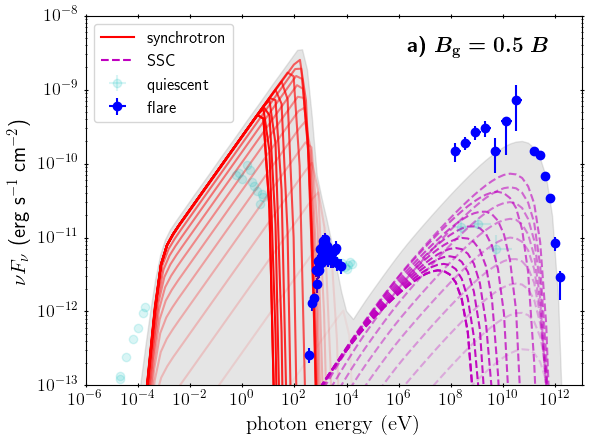}
    \caption{SED showing the effect of increasing the guide field to be half of the total magnetic field, using the best fit parameters for Fig. \ref{fig:BF} which are listed in Table \ref{tab:BF}. Fainter lines indicate earlier times in the time evolution of the SED. It can be seen that the synchrotron emission is slightly diminished relative to the best fit case from Fig. \ref{fig:BF}a (grey silhouette), but there is greater reduction in the SSC emission. This is because as the guide field component is larger than the best fit case, the total magnetic field strength remains relatively higher, causing higher radiative losses. The total magnetic energy is therefore more quickly depleted, giving a smaller final blob size of $8 \times 10^{13}~\rm{m}$. From Eqn. \ref{eq:PproptoR}, less SSC emission is produced.}
    \label{fig:BgSED}
\end{figure}

\subsection{Flaring Profiles}\label{sec:flareProf}
Fig. \ref{fig:BF}b illustrates our best fit to the 2016 TeV flare of BL Lacertae. The entire flare takes place over the course of $\approx~2$ hours, and our model is able to provide a reasonable fit to the data with a $\chi_{\nu}^2 = 1.95$. The radiative emission from our reconnecting plasmoid well replicates the slow rise and rapid decline of the flare. This is important because it demonstrates that a reconnecting plasmoid is able to produce powerful flares on short timescales. 

Fig. \ref{fig:BF}c shows that the entire light curve illustrated in Fig. \ref{fig:BF}b occurs in the phase where the emitting plasmoid becomes particle dominated with $\sigma < 1$. This can be used to explain the characteristic slow rise and rapid decay. During the rise of the flare, the radius of the plasmoid is increasing, thus increasing its synchrotron emissivity. From Eqn. \ref{eq:tm}, the increase in $R$ also increases the value of $\tau_{\rm{merge}}$, making the acceleration timescale longer. This affect alone would cause the magnitude of the flare to increase less rapidly over time, however in the $\sigma<0$ case an additional effect becomes more important. Fig.\ref{fig:BF}a illustrates that SSC emission from later time SEDs have a higher maximum energy, because of the decline in $B$ shown throughout the emission time in Fig. \ref{fig:BF}c. The maximum electron energy depends on a balance between radiative losses, which decrease with decreasing $B$, and the rate of particle acceleration, which decreases with increasing $R$. In this case, the decline in $B$ is much more rapid than the growth of $R$, and so particle acceleration can reach higher energies, causing the flare to rise. The rapid decline is caused by the same property that prevents Compton-dominant SEDs from this model. Namely, $B$ rapidly falls as $\sigma$ approaches zero, quickly depleting the synchrotron seed photons needed for SSC emission. In the VHE band plotted in Fig. \ref{fig:BF}b, the radiative lifetime for the high energy electrons radiatively emitting in that band is very small, providing the dramatic decline.

The inclusion of a guide field effectively sets an upper limit to the minimum magnetic field. This in turn limits the minimum degree of magnetisation the reconnecting plasmoid can have when it is assumed to leave the reconnection layer. One consequence of this is that once the magnetic energy associated with $B_{\rm{rec}}$ has been completely transferred to particles which are accelerated, the electron population evolves in a magnetic field equal to the guide field. As $B_{\rm{g}}$ is constant, the radiative lifetimes of electrons at different energies are now fixed. Eqns. \ref{Psync} and \ref{P_IC} indicate that the highest energy electrons radiate most quickly. Accordingly, the evolution of the flare profile once $B_{\rm{rec}}$ has been depleted depends on this behaviour, with the decline of the flare in this regime showing an exponential decay dependent on the shortest radiative lifetime, namely $F_{\nu} \sim \exp(-t/\tau_{\rm{rad}}(E, t))$. This is because over time TeV emission from these plasmoids becomes progressively dominated by lower energy electrons, until it ceases to be emitted altogether when the maximum electron energy is too low to produce TeV photons via SSC.

Whilst particle dominated flares typically have rapid decay times, the model is also capable of producing flares which are close to being symmetric, as indicated by Fig. \ref{magflare}. In this scenario, the rise time of the flare is governed by the merging timescale, which Eqn. \ref{eq:tm} shows decreases with increasing $R$. In this case, $\sigma > 1$ for the duration of the emission period, thus there is little change on the value of $B$ and the radial growth has the most influence on the flare profile. The increasing value of $\tau_{\rm{merge}}$ steadily reduces the acceleration efficiency, until a peak is reached which coincides with where the electron radiative losses are balanced with the particle acceleration. After this point, as $R$ keeps increasing, the flare begins to decline. In the rest frame of the emitting plasmoid, the rate of decline of the flare decreases as the peak frequency of the flare decreases and the emitting electrons decrease in energy and have progressively longer radiative lifetimes, with the emitted flux eventually becoming negligible. This gives a slower decay time relative to the rise time in the plasmoid rest frame. However, if the plasmoid is accelerating as was the case in Fig. \ref{magflare}, the doppler factor is higher at late times reducing the time interval. This effect makes the flare look more symmetric to an observer. If the emitting plasmoid has reached its peak velocity, a longer decline than rise may instead be observed. This result has previously been found in the literature in radio data for extragalactic radio sources \citep{1999ApJS..120...95V,2016MNRAS.460.1790G}, though this has not before been explicitly predicted by theory. Such a flare profile was typical of magnetically dominated plasmoids, and it should be noted that the parameters in Table \ref{tab:BF} corresponding to Fig. \ref{magflare} are included for completeness only. We find that for flares at different peak frequencies originating from both magnetically and particle dominated regions that the flaring profiles are qualitatively the same as for the TeV band, with the only difference being that the characteristic rise and decay times reflect the radiative lifetime of the emitting electrons.

\subsection{Implications for PIC Simulations}\label{sec:PIC}

\begin{figure}
	\includegraphics[width=\columnwidth]{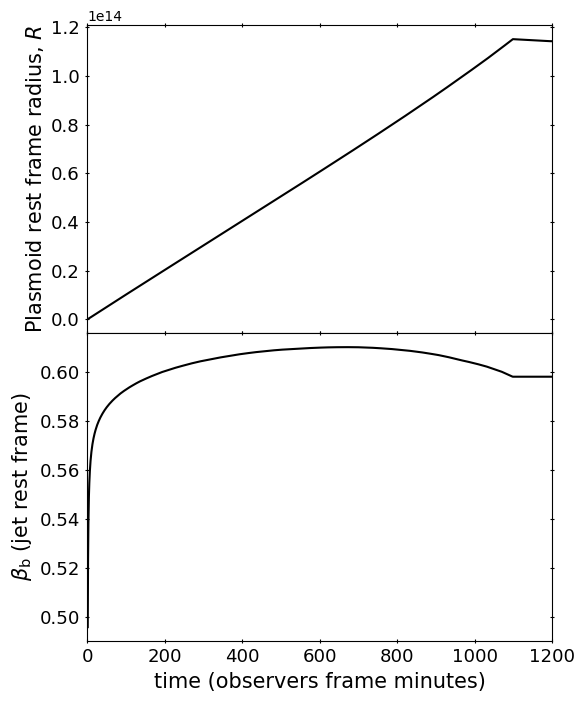}
    \caption{Upper: The radial growth of the plasmoid required to fit to the TeV flare of BL Lacertae with free parameters given in Table \ref{tab:BF} had a radial growth profile which was approximately linear. Lower: $\beta_{\rm{b}}$ profile \citep{2016MNRAS.462...48S} of the plasmoid uses to fit the SED and TeV light curve in Fig. \ref{fig:BF}. It can be seen that the velocity initially increases, peaking at around $0.6c$, before declining before the end of the simulation. When the magnetic energy in the reconnecting field has been completely depleted, it is assumed the emitting plasmoid leaves the reconnection layer and no longer merges. After this point, it is assumed to travel at a constant velocity and its radius contracts slightly due to radiative emission, and this occurs in the figure for $t \geq 1100~\rm{mins}$.}
    \label{fig:RLvel}
\end{figure}

One difference between our model and PIC simulations is that the final power law on our electron population does not directly depend on $\sigma$. This is a direct consequence of there being no $\sigma$ dependence in Eqn. \ref{DLeqn} and is a limitation of our model. Work undertaken by \citet{2018ApJ...862...80B} find that the spectral index on the electron population is $p \approx 1.8 + 0.7/\sqrt{\sigma}$. Our power law index is given in Eqn. \ref{SSDL} as $p = 1+\tau_{\rm{merge}}/\left(\alpha \tau_{\rm{esc}}\right)$, and converges to $\approx 1$ under our assumption that the emitting electrons were trapped in the reconnection layer with $\tau_{\rm{esc}} = \infty$. We chose this because it allowed for the reduction in free parameters and simplified the model, and our main conclusions regarding the Compton dominance of the SEDs have negligible dependence on this assumption. 

One possibility to reconcile our electron population with PIC results would be to change the value of $\tau_{\rm{esc}}$ to depend on the size of the reconnection layer, $L$. Taking $\tau_{\rm{esc}} \approx L/c $ \citep{2014PhRvL.113o5005G} would likely mean that our electron spectral index converges to $\approx 1$ as $L$ needs to be large for astrophysical objects. Another option could be that the efficiency of the particle acceleration process, which we quantify with $\alpha$, is not constant as assumed but depends on the magnetisation. Therefore, whilst we do not expect this to alter the main conclusions of our work, we aim to investigate further in future work what the impact of including $\sigma$ in the acceleration or escape terms would be. 

Throughout this work, we have adopted the velocity profile of \citet{2016MNRAS.462...48S}. Our success in fitting to the timescales of the TeV flare in Fig. \ref{fig:BF} suggest that this profile works very well, and is capable of producing powerful flares on short time-scales. The velocity of the best fit emitting plasmoid as it travels in the reconnection layer is shown in Fig. \ref{fig:RLvel}. The figure shows that this particular plasmoid only ever achieved mildly relativistic speeds of $\beta \approx 0.6~c$, and after some rapid initial acceleration reached a peak velocity before slowing slightly beyond $t \geq 700~\rm{mins}$. This may occur because we are in a much higher $\sigma$ regime than for which Eqn. \ref{blobvel} was derived \citep[see][]{2016MNRAS.462...48S} and extrapolate the use of this formula to an environment for which it may not be valid.

\subsection{Can the Reconnection Layer Physically Exist Within the Jet?}

The final size of the emitting plasmoid in the reconnection layer from our fit to the 2016 TeV flare of BL Lacertae was $R_{\rm{f}}=1.2 \times 10^{14}~\rm{m}$. Furthermore, the plasmoid travelled a distance $L = 9.6 \times 10^{14}~\rm{m}$ in the rest frame of the jet. These are comparable to the typical radii used to model the quiescent emission from blazar jets in single zone emission models \citep{2002A&A...386..833G}, which implies that the reconnecting plasmoid needs to occupy a substantial fraction of the entire jet. We consider the implications of this large size with respect to location within the jet. 

First, we consider whether the emitting plasmoid model can explain the $\approx 2$ hour TeV flare of BL Lacertae whilst satisfying light travel time arguments. Throughout the emission of the 2016 TeV flare depicted in Fig. \ref{fig:BF}, the observed Doppler factor was around $\delta_{\rm{p}} \approx 72$, although this changed slightly with the variation in $v_{\rm{blob}}$. Substitution of the final plasmoid radius and $\delta_{\rm{p}}$ into Eqn. \ref{size} constrains the permitted variability timescale to $t_{\rm{var}} \geq 100~\rm{min}$, thus our best fit is not prohibited when considering the light crossing time.

Secondly, we constrain the location of the reconnection region in the blazar jet by assuming it to be in pressure equilibrium with the surrounding plasma. To do this, we assume a jet which conserves total magnetic energy as it expands radially, such that the radial dependence on the magnetic field of $B_{\rm{jet}} \propto R^{-1}$ \citep{PC:2012}. By assuming the jet to be in equipartition, an estimate of the jet radius at the point of pressure equilibrium can be found. We take the power for the jet in BL Lacertae in the observers frame as $P_{\rm{jet}} = 9.0\times 10^{36}~\rm{W}$ \citep{2013MNRAS.436..304P}. In such a model, this power is contained in a conical section of the jet of length one light second, $l_{\rm{c}}$, giving an observers frame energy density of,

\begin{equation}
U_{\rm{jet}} \approx \frac{P_{\rm{jet}}}{ \pi R_{\rm{jet}}^2 l_{\rm{c}}}.
\label{eq:ujet}
\end{equation}
We are interested in the frame in which the jet is at rest. The energy density of the jet in this frame, $U_{\rm{jet}}'$ is related to, $U_{\rm{jet}}$ by \citep[e.g.][]{PC:2012},

\begin{equation}
U_{\rm{jet}}' \approx \frac{U_{\rm{jet}}}{ \Gamma_{\rm{j}}^2}.
\label{eq:ujetrf}
\end{equation}
Table \ref{tab:BF} gives the total energy density of the reconnection plasmoid in its rest frame as $U_{\rm{TOT}} = 10^{-4}~\rm{J m}^{-3}$. To compare to the jet energy density, we convert this to the jet frame as $U_{\rm{TOT}}' \approx \Gamma_{\rm{b}}^2 U_{\rm{TOT}}$ by analogy to Eqn. \ref{eq:ujetrf}, where $\Gamma_{\rm{b}} \approx 1.25$ using $v_{\rm{b}} \approx 0.6~c$ as the approximate peak velocity from the lower panel in Fig. ref{fig:RLvel}. Combining Eqns. \ref{eq:ujet} and \ref{eq:ujetrf} and assuming pressure equilibrium such that $U_{\rm{jet}}' = U_{\rm{TOT}}'$ allows us to solve for the jet radius as,
\begin{equation}
R_{\rm{jet}}' = \sqrt{ \frac{P_{\rm{jet}}}{\Gamma_{\rm{j}}^2 \pi l_{\rm{c}} U_{\rm{TOT}}' } },
\end{equation}
yielding $R_{\rm{jet}}' \approx 3 \times 10^{14}~\rm{m}$ as the radius of the jet at the location of the plasmoid in the plasmoid rest frame. This is only a factor $\approx 2$ larger than the radius, and smaller than estimated length of the reconnection layer of $L =   9.6 \times 10^{14}~\rm{m}$ calculated using Eqn. \ref{eq:L}. This estimated jet radius is likely to be problematic when considering the distance the plasmoid has to travel to grow large enough to be observable, it seems implausible to us that an SSC emitting reconnecting plasmoid can explain the BL Lacertae flare since the size of the reconnecting region and plasmoid are both comparable to the size of the entire jet radius. Therefore, this plasmoid is unlikely to physically reside in the jet.

One possible explanation for the required large plasmoid size comes from a limitation of our model where we have neglected emission from external Compton radiation. The inclusion of this may result in smaller plasmoids which are able to emit a larger proportion of high energy radiation than a similarly sized emitting plasmoid only radiating via SSC. Inclusion of this may be able to produce more realistic reconnection plasmoids which are not synchrotron dominated and are small compared to the jet radius. This is something we aim to investigate in future work.

\section{Summary and Conclusions}\label{sec:conc}

We have presented a macroscopic emission model which tracks the growth and velocity of a radiating plasmoid as it travels through a reconnection layer in a blazar jet to assess the feasibility of magnetic reconnection powering blazar flares. Our leptonic model accounts for particle acceleration within the reconnection layer and computes the radiative emission from the reconnecting plasmoid, including synchrotron and synchrotron self-Compton emission. As an example we simultaneously fit the SED and TeV light curve for the 2016 TeV flare of BL Lacerate. The main conclusions of this work may be summarised as:

\begin{itemize}
  \item Our reconnection model produces synchrotron-dominated flares, and cannot produce Compton-dominated flares. This means that the plasmoid model is not able to fit the 2016 TeV flare of BL Lacertae because synchrotron emission is overproduced relative to SSC gamma-rays.
  \item Reconnecting plasmoids are able to produce powerful, rapid TeV flares. The model is able to fit well to the observed TeV light curve of the 2016 flare, the first time a reconnection emission model has been able to fit to a TeV blazar time series.
  \item From an extensive parameter search we find that reconnecting plasmoids can produce a variety of lightcurve shapes (fast rise slow-decay, symmetric and slow rise fast-decay) depending primarily on the magnetisation of the plasma and the merge timescale of plasmoids.
  \item The final size of emitting plasmoids which are luminous enough to be detected is problematically large, they are comparable to the estimated jet radius. This is similar to radii used to model quiescent jet emission in one-zone jet models and calls into question the physical viability of reconnection powering such a flare if the emission mechanism is SSC.
\end{itemize}
 
In this work we have only considered synchroton and SSC emission from the reconnecting plasmoid. In the future we will investigate the effect of including inverse-Compton scattering of external photons which may help to alleviate the problems of overproduction of synchrotron emission and the problematically large size of observable plasmoids.

\section*{Acknowledgements}

We thank our anonymous referee for constructive comments that have improved the manuscript. PJM acknowledges support from a Hintze Scholarship. WJP is supported by a Junior Research Fellowship from University College, University of Oxford. GC acknowledges support from STFC grants ST/N000919/1 and ST/M00757X/1 and from Exeter College, Oxford. This work was supported by the Oxford Hintze Centre for Astrophysical Surveys which is funded through generous support from the Hintze Family Charitable Foundation. 




\bibliographystyle{mnras}
\bibliography{General}




\bsp	
\label{lastpage}
\end{document}